\documentclass[aps,floatfix,groupedaddress,nofootinbib,superscriptaddress,twocolumn]{aastex63}
\usepackage[T1]{fontenc}
\usepackage{microtype}
\usepackage{afterpage,amsmath,amssymb,graphicx,natbib, color}
\usepackage{hyperref}
\usepackage[english]{babel}
\usepackage{blindtext}
\usepackage{csquotes}
\usepackage[utf8]{inputenc}
\usepackage{enumitem}
\usepackage{graphicx}
\usepackage{hyperref}
\usepackage{mathtools}
\usepackage{comment}
\usepackage{amsmath}
\usepackage{ amssymb }
\usepackage{lipsum} 
\usepackage{float}
\usepackage{cleveref}
\usepackage[caption=false]{subfig}
\usepackage{placeins}
\usepackage{xcolor}
\usepackage{multirow}
\usepackage{float}
\usepackage{makecell}

\newcommand{\be}{\begin{equation}}
\newcommand{\ee}{\end{equation}}
\newcommand{\bfig}{\begin{figure}}
\newcommand{\efig}{\end{figure}}

\newcommand{\limpy}{\texttt{LIMpy}}

\begin{document}
\title{Cross-correlation Techniques to Mitigate the Interloper Contamination for Line Intensity Mapping Experiments}

\correspondingauthor{Anirban Roy}
\email{ar689@cornell.edu}

\author{Anirban Roy}
\affiliation{Department of Astronomy, Cornell University, Ithaca, NY 14853, USA}
\affiliation{Department of Physics, New York University, 726 Broadway, New York, NY, 10003, USA}  
\affiliation{Center for Computational Astrophysics, Flatiron Institute, New York, NY 10010, USA} 

\author{Nicholas Battaglia}
\affiliation{Department of Astronomy, Cornell University, Ithaca, NY 14853, USA}

\begin{abstract}
Line intensity mapping (LIM) serves as a potent probe in astrophysics, relying on the statistical analysis of integrated spectral line emissions originating from distant star-forming galaxies. While LIM observations hold the promise of achieving a broad spectrum of scientific objectives, a significant hurdle for future experiments lies in distinguishing the targeted spectral line emitted at a specific redshift from undesired line emissions originating at different redshifts. The presence of these interloping lines poses a challenge to the accuracy of cosmological analyses. In this study, we introduce a novel approach to quantify line-line cross-correlations (LIM-LLX), enabling us to investigate the target signal amidst instrumental noise and interloping emissions. For example, at a redshift of approximately $z\sim3.7$, we observed that the measured auto-power spectrum of CII\,158 exhibited substantial bias, from interloping line emission. However, cross-correlating CII\,158 with CO\,(6-5) lines using a FYST-like experiment yielded a promising result, with a Signal-to-noise ratio (SNR) of $\sim 10$. This measurement is notably unbiased. Additionally, we explore the extensive capabilities of cross-correlation by leveraging various CO transitions to probe the tomographic Universe at lower redshifts through LIM-LLX. We further demonstrate that incorporating low-frequency channels, such as 90 GHz and 150 GHz, into FYST's EoR-Spec-like experiment can maximize the potential for cross-correlation studies, effectively reducing the bias introduced by instrumental noise and interlopers.
\end{abstract}


\section{Introduction} 
Line Intensity Mapping (LIM) is a promising approach that allows us to probe the three-dimensional (3D) structure of the Universe beyond the traditional galaxy-by-galaxy surveys and explore the collective properties of atomic and molecular emissions from the interstellar medium (ISM) of galaxies \citep{Visbal2010, Visbal2011, Kovetz2017LIM_report, Bernal2022-sfr}. By indirectly tracing the cumulative radiation from multiple atomic and molecular emission lines, LIM offers a unique statistical view of large-scale structures, providing valuable insights into galaxy formation and evolution across cosmic time \citep{Limfast2, Zhang-galaxy-property}. This technique has emerged as a tool for investigating the cosmic landscape across vast cosmic volumes and exploring the evolution of galaxies, intergalactic gas, and cosmic star formation \citep{Bernal2022-sfr, Zhou-lim-sfr}. The exploration of several lines, including fine-structure emissions like [CII] (157.7 $\mu \text{m}$) and [OIII] (52 \& 88.4 $\mu \text{m}$), along with rotational emission lines from CO, garners substantial interest in upcoming LIM experiments conducted at millimeter and sub-millimeter wavelengths \citep{Suginohara1998, Righi2008b, Lidz2011_CO, Carilli2011, Fonseca:2016, Gong2017, Kovetz2017LIM_report, Chung2018CII, PadmanabhanCO, Padmanabhan_CII, Dumitru2018, Chung2018CO, Kannan:2021ucy, Murmu:2021ljb, Karoumpis2021, Limfast2, Slick-Garcia}. These lines are of particular interest due to their inherent brightness properties and frequency overlap with the current, millimeter and sub-millimeter wave LIM experiments. These experiments, including the EoR-Spec instrument at FYST\footnote{\href{https://www.ccatobservatory.org/}{https://www.ccatobservatory.org/}} \citep{CCAT-prime2021}, SPHEREx\footnote{\href{https://spherex.caltech.edu/}{https://spherex.caltech.edu/}} \citep{SPHEREx-science-paper2018}, TIME \citep{Time-science-2014}, CONCERTO\footnote{\href{https://mission.lam.fr/concerto/}{https://mission.lam.fr/concerto/}} \citep{CONCERTO-science-2020}, COMAP\footnote{\href{https://comap.caltech.edu/}{https://comap.caltech.edu/}} \citep{COmap-science-2021}, and EXCLAIM \citep{EXCLAIM-2020}, have been designed to observe several atomic and molecular lines spanning various redshifts.

Among the various aspects of LIM observations, Line-Line Cross-Correlation (LIM-LLX) emerges as a promising avenue for gaining insight into the astrophysics driving galaxy formation. This is achieved through the detection of cross-correlated lines originating from the same sources. Rather than analyzing individual lines, this cross-correlation approach capitalizes on the combined power of multiple lines, revealing their correlations and connections between different astrophysical processes. By targeting several emission lines simultaneously at a particular redshift, LIM-LLX could unlock novel perspectives on cosmic star formation rate (SFR), large-scale structure growth, and the cosmic distribution of neutral hydrogen \citep{Kovetz2017LIM_report, Karkare:2018sar, Bernal:2019gfq, Silva:2019hsh}. These far-reaching insights hold the potential to deepen our understanding of cosmic phenomena, uncovering the complex interplay between galaxies and the cosmic environment. Furthermore, cross-correlation studies between LIM and various tracers, including secondary CMB anisotropies, galaxy surveys, and the cosmic Infrared background, are crucial for resolving degeneracies between astrophysical and cosmological parameters \citep{Schaan:2021gzb, Zhou-lim-sfr, Maniyar-lensing-null, Fronenberg2023}. 
For example, cross-correlations between LIM and 21cm maps can be instrumental in recovering the auto-correlations within the 21cm maps, particularly through the application of estimators like the Least Squares Estimator \citep{Beane2018, Beane2019, McBride2023}.

Limiting the potential of LIM-LLX, is the presence of interloper contamination \citep{Lidz2016-interloper, Gong2020}, which can bias the overall signal. Interloper contamination arises from foreground and background emissions unrelated to the targeted emission lines.
Properly mitigating interloper contamination is, therefore, essential for ensuring the fidelity of the detection LIM signals and extracting accurate astrophysical information from the observed data.

In this work, we explore the critical role played by interloper contamination in LIM auto power spectra and LIM-LLX studies. We begin by elucidating the theoretical framework of LIM, highlighting its capacity to probe cosmic structures through the collective emission from the ensembles of galaxies. A significant portion of this article is dedicated to comprehensively understanding interloper contamination and its effect on signal detection in the presence of instrumental noise and beam smearing effect. Afterwards, we delve into the simulation-based approach to explore the potential of LIM-LLX, emphasizing its ability to tap into the synergistic potential of multiple emission lines to reduce the bias on signal detection. We analyze the multifaceted sources of contamination, taking into account the effects of instrumental noise and interlopers modeling. As we navigate through the complexities of noise and interloper-induced biases, we explore the methodologies developed to identify and extract target line correlations amidst the contaminating signals. Moreover, we discuss the impact of interloper contamination on LIM-LLX measurements and the challenges associated with instrumental and observational limitations. By examining current state-of-the-art techniques for interloper mitigation, we underscore the need for robust error estimation and data analysis methodologies to secure reliable and accurate interpretation of LIM observations.

The structure of this paper is outlined as follows: In Section \ref{sec:modelling}, we describe the astrophysical modeling of various LIM signals across a broad redshift range. This includes the identification and characterization of interlopers and the presence of instrumental noise along with a brief overview of the fundamental concept of intensity mapping. In Section \ref{sec:quantifying}, we embark on the impact of interlopers on both the auto and cross-power spectrum of target lines. To quantify their influence, we introduce a bias parameter, shedding light on how it affects the detection of the desired signal. Moving forward to Section \ref{sec:detectability}, we focus towards the forecasting of SNR and the overall efficacy of LIM-LLX techniques. This analysis extends across a wide range of redshifts, allowing us to gain insights into the method's performance for different pairs of line-line cross-correlations. Finally, in Section \ref{sec:conclusion} we consolidate our findings and present concluding remarks, elucidating the relevance of LIM-LLX studies in the context of future LIM experiments. 
In this paper, we define the desired signal at a particular redshift as the \enquote{target} signal, which we aim to probe through LIM experiments. Additionally, we define the \enquote{measured} signal as the desired signal that will ultimately be detected after mitigating the bias due to noise and interlopers.

\begin{figure}[h!]
\includegraphics[width=0.5\textwidth]{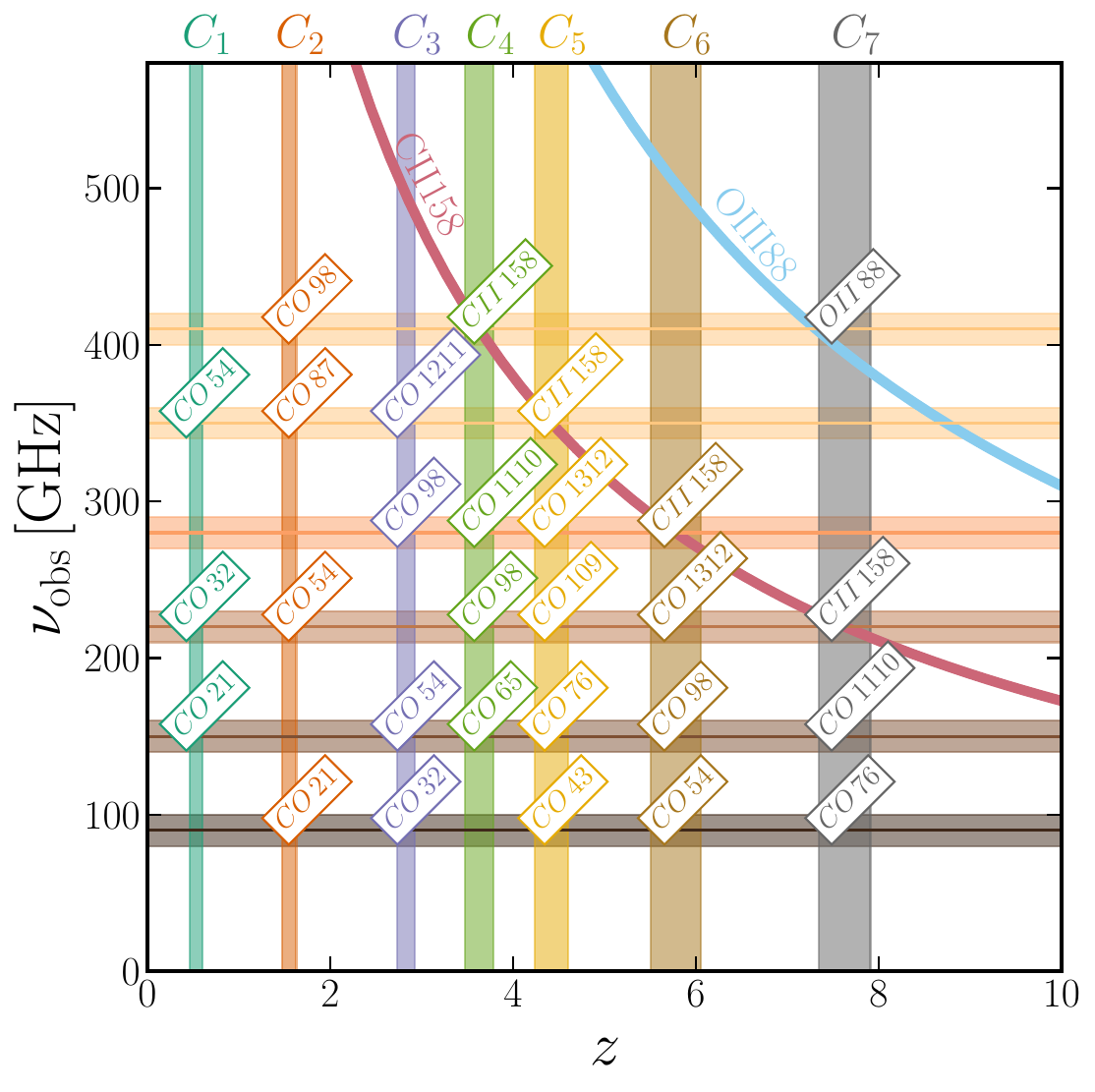} 
\caption{We demonstrate the numerous cross-correlation opportunities between distinct molecular lines (CO) and atomic emission lines (CII\,158 and OIII\,88). The horizontal lines on the plot represent the frequency channels, while the shaded areas around them denote their corresponding bandwidths. Furthermore, the vertical lines illustrate the various groups of lines suitable for cross-correlation analysis, all originating from roughly the same redshift. Therefore, LIM-LLX across a wide redshift range provides a tomographic perspective, revealing the large-scale cosmic structure and enabling the exploration of cosmic history and the evolution of galaxies.}
\label{fig:nu_z}
\end{figure}

Throughout this work, we adopt the cosmological framework of a flat $\Lambda$CDM universe, described by the cosmological parameters determined under the Planck TT, TE, EE+lowE+lensing findings \citep{P18:main}. In the subsequent sections of this paper, we represent atomic line emissions by combining the line name with its wavelength in micrometres, such as CII\,158. For molecular line emissions originating from CO, we employ a nomenclature that specifies the upper rotational transition level to the lower level, such as CO (1-0). This naming convention aligns with the same convention practices in LIMpy \citep{limpy}.

\section{Modelling of LIM-LLX}\label{sec:modelling}
This section presents an overview of the methods employed to generate LIM signals that could be targeted by LIM experiments, like EoR-Spec on FYST. Our primary objective is to provide a detailed, step-by-step description of the procedures utilized to create simulated LIM data. These simulations are created to test the fidelity of analysis methods before observational data is available. 

\begin{table}
\centering
\begin{tabular}{cccc}

Group & Redshift & Line Names & Frequency (GHz) \\
\hline
C1 &  $0.58$ & \textbf{CO\,(3-2)}  & \textbf{220}  \\
 &   & \textbf{CO\,(5-4)}  & \textbf{350}  \\
 &  & CO\,(2-1) & 150 \\
 \hline
 
C2 & 1.6  & \textbf{CO\,(2-1)} &  \textbf{90}  \\
 &   &  \textbf{CO\,(5-4)} & \textbf{220} \\
 &   & CO\,(8-7)  &  350 \\
  &   & CO\,(9-8)  & 410  \\

\hline
C2$^\prime$ & 1.6  &  \textbf{CO\,(5-4)} & \textbf{220} \\
 &   & \textbf{CO\,(8-7)}  &  \textbf{350} \\
\hline

C3 & 2.9  & \textbf{CO\,(3-2)}  &  \textbf{90} \\
    &   &  \textbf{CO\,(5-4)}  & \textbf{150}  \\
     &   & CO\,(9-8)  &  280 \\
      &   & CO\,(12-11)  &  350 \\
\hline

C3$^\prime$     & 2.9  & \textbf{CO\,(9-8)}  &  \textbf{280} \\
      &   & \textbf{CO\,(12-11)}  &  \textbf{350} \\
\hline

C4  & 3.71  & \textbf{CII\,158}  & \textbf{410}  \\
 &   & \textbf{CO\,(6-5)}  & \textbf{150} \\
    &   &  CO\,(11-10)  & 280  \\
     &   &  CO\,(9-8)  & 220  \\
    
\hline
C4$^\prime$  & 3.71  & \textbf{CII\,158}  & \textbf{410}  \\
    &   &  \textbf{CO\,(9-8)}  & \textbf{220}  \\
\hline

C5 & 4.4 & \textbf{ CII\,158} & \textbf{350} \\
&   & \textbf{CO\,(4-3)}  & \textbf{90} \\
&   & CO\,(7-6)  &  150 \\
&   & CO\,(10-9)  & 220 \\
&   & CO\,(13-12)  & 280  \\
\hline

C5$^\prime$ & 4.4 & \textbf{ CII\,158} & \textbf{350} \\
&   & \textbf{CO\,(10-9)}  & \textbf{220} \\
\hline
C6 & 5.85 & \textbf{CII\,158} & \textbf{280} \\
 &  & \textbf{CO\,(5-4)}  & \textbf{90} \\
    &  & CO\,(13-12)  & 220 \\
    &  & CO\,(9-8)  & 150 \\
\hline
C6$^\prime$ & 5.85 & \textbf{CII\,158} & \textbf{280} \\
    &  & \textbf{CO\,(13-12)}  & \textbf{220} \\
\hline

C7 & 7.6 & \textbf{CII\,158} & \textbf{220}  \\
&  &  \textbf{OIII\,88} & \textbf{410} \\
&  & CO\,(11-10) & 150 \\
&  & CO\,(7-6) &  90\\
\hline
\end{tabular}
\caption{We introduce cross-correlation groups, line names sharing a common redshift, the cross-correlation redshifts, and the associated observation frequency channels. Bold line names highlight the brightest lines within each group, which we use for cross-correlation analyses in this paper, along with their corresponding frequency channels. We highlight groups marked with prime signs, as these particular lines fall within the frequency coverage of FYST, while non-prime cross-correlation groups make use of the 90 and 150 GHz frequency channels. \\}
\label{tab:cross_groups}
\end{table} 

In Figure \ref{fig:nu_z}, we illustrate the capability of LIM-LLX to probe several lines originating from nearly the same redshift. For this analysis, we utilize four frequencies ranging from 220 to 410 GHz, consistent with the frequency coverage described in the specifications of the EoR-Spec instrument on FYST \citep{CCAT-prime2021}. To maximize the utility of line-line cross-correlation, we incorporate two additional frequency channels at 90 and 150 GHz. The inclusion of these channels increases the ability to detect LIM-LLX with CII\,158 lines at high redshift and also enables the cross-correlation of different CO J-level transitions, providing insights into astrophysical phenomena at low redshifts. We label the cross-correlation groups C1 to C7 in ascending order in redshift and for every group we can cross-correlate different frequency channels to pick up the LIM-LLX signal which is coming from the same sources (or redshift range). This figure emphasizes the significance of measuring LIM-LLX across a wide redshift range and the potential of cross-correlation techniques in mapping the Universe in a tomographic systematic way. 

In Table \ref{tab:cross_groups}, we present the possible cross-correlations between several lines by using six frequencies ranging from 90 to 410, and the redshift of cross-correlations. For the sake of completeness, we listed all the higher-order J-level transitions of CO molecules belonging to a particular group that can be potentially cross-correlated. Although we have calculated all possible cross-correlations, we only report the brightest cross-correlation in the result section, these are bolded in the table. We adopt the configuration of an EoR-Spec-like instrument on FYST for making the forecasts. However, any other experiments can roughly scale the forecasts based on their experimental configurations. We note that the different lines belonging to the same group do not have \textit{exactly} the same redshift but they overlap with each other if they fall under a broad frequency coverage around the central frequencies of a channel for EoR-Spec. 

We provide a summary of crucial parameters, including white noise, frequency resolution, and beam size, curated for six distinct frequency channels in Table \ref{tab:obs}. The quantities for frequency channels spanning from 220 GHz to 410 GHz have been sourced directly from FYST's science book \cite{CCAT-prime2021}, determined based on the telescope's configuration. For the additional channels operating at 90 GHz and 150 GHz, we scale these parameter values from their higher-frequency counterparts, following the scaling in \cite{CCAT-prime2021} for the broadband camera. While this method serves as a reasonable approximation for estimating white noise and beam size, it is important to note that the precise values of these parameters are contingent on the specific design of the telescope. 

\subsection{Theoretical Background}
The expression for the intensity of emission lines at redshift $z_{\rm em}$ can be formulated as follows:

\begin{equation}
I_{\rm line}(z)= \frac{c}{4\pi}\frac{1}{{\nu_{\rm rest}}H(z_{\rm em})} \int_{M_{\rm min}}^{M_{\rm max}} L_{\rm line}(M,z)\frac{dn}{dM}dM.
\end{equation}\label{eq:Iline}

Here, the symbol $c$ represents the speed of light in a vacuum, and $H(z_{\rm em})$ stands for the Hubble parameter at the redshift of the emitted line. The function $dn/dM$ corresponds to the halo mass distribution. The variables $M_{\rm min}$ and $M_{\rm max}$ indicate the minimum and maximum halo masses contributing to the intensity maps, respectively. We denote the luminosity of several lines by $L_{\rm line}$ which we model as a function of halo mass and redshift. For this paper, we implement various models from the $\limpy$ package which is described in detail in \citet{limpy}.

Simulations for LIM have the advantage that they need not resolve every individual source, only the sources that contribute significantly to $I_{\rm line}(z)$. This advantage allows one to run simulations that resolve a minimum mass halo, ultimately leading to a reduction in simulation time due to the reduction in halo mass resolution required. Throughout these simulations, we store all the intensity grids at various redshifts, essential for the subsequent calculations for auto and cross-power spectrum.

The 3D line intensity auto-power spectrum within the simulation box can be expressed as:

\begin{equation}
\Delta^2_{\rm line}(k) = \frac{1}{V_{\rm box}} \frac{k^3}{2\pi^2} \langle \tilde{I}^2(k)\rangle.
\label{eq:pk_sims}
\end{equation}
In this equation, $V_{\rm box}$ denotes the total volume of the simulation box, and $\tilde{I}$ represents the Fourier transform of the simulated line intensity projected onto a 3D grid cell (voxel), $I_{\rm cell}$. To calculate $\tilde{I}$, we utilize the NumPy FFT module \citep{numpy}.

The intensity of each grid cell can be written as \citep{Dumitru2018}:

\begin{equation}
I_{\rm cell} = \frac{c}{4\pi} \frac{1}{\nu_{\rm rest} H(z)} \frac{L_{\rm line, cell}}{V_{\rm cell}}\,.
\end{equation}
Here, $V_{\rm cell}$ represents the volume of the cell. Similarly, we calculate the cross-power spectrum from the simulation box using the equation \citep{Dumitru2018}:

\begin{equation}
\Delta^2_{\rm line, x}(k) = \frac{1}{V_{\rm box}} \frac{k^3}{2\pi^2} \frac{\langle \tilde{I_1}(k)\tilde{I_2}^{*}(k) + \tilde{I_1}^{*}(k)\tilde{I_2}(k) \rangle}{2}\,.
\label{eq:pk_sims_cross}
\end{equation}
Here, $I_1$ and $I_2$ refer to the intensities of the first and second lines that we want to cross-correlate. The Fourier transform of the intensity grid and its complex conjugate are denoted as $\tilde{I}$ and $\tilde{I}^{*}$, respectively. 
The equation above can be used to calculate the cross-correlation between any two fields, not limited to line-line cross-correlations.

\begin{figure*}[t]
\includegraphics[width=\textwidth]{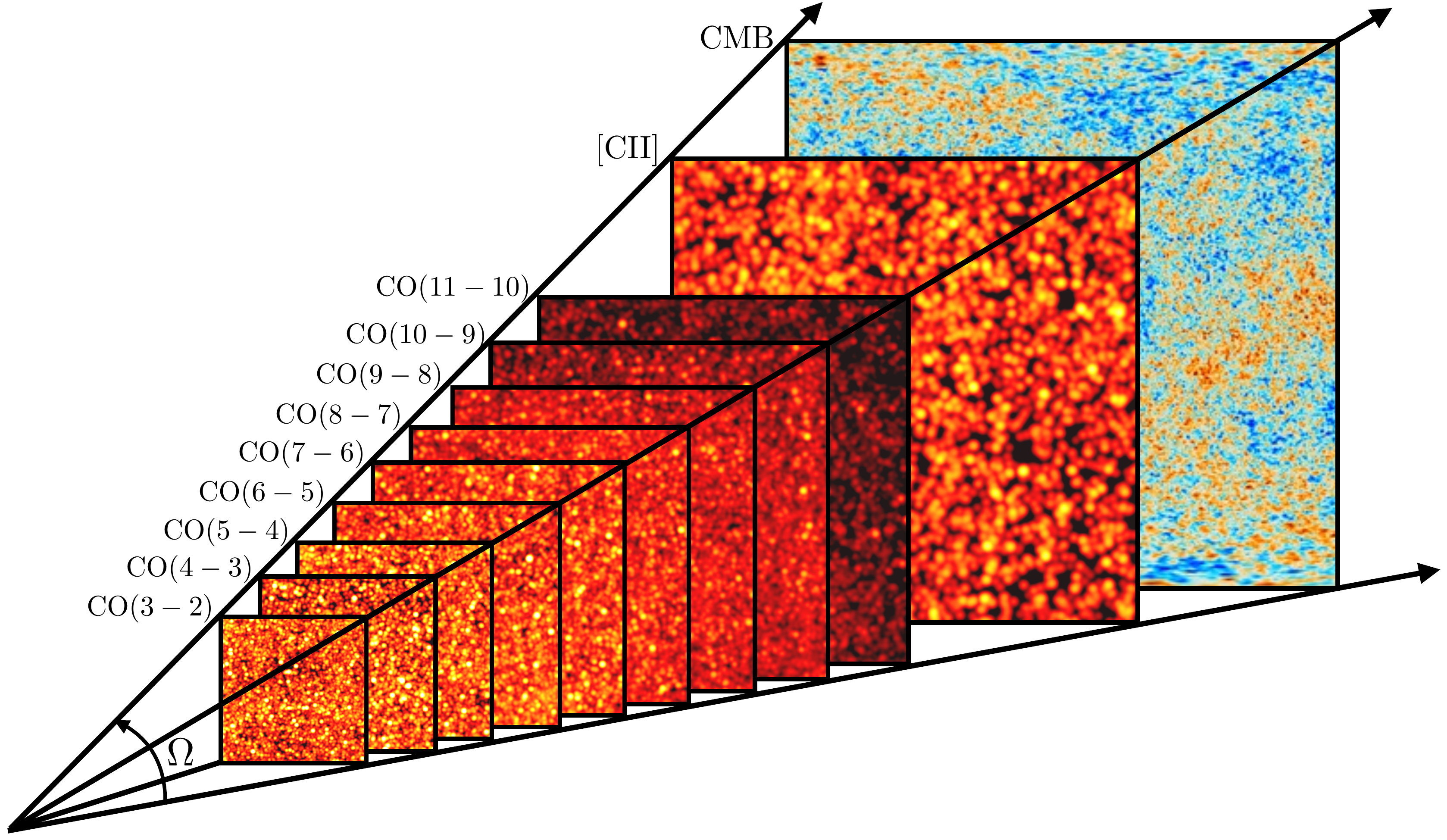} 
\caption{A lightcone visualization within a solid angle $\Omega$, featuring panels depicting both the signal and interlopers, with the Cosmic Microwave Background (CMB) serving as a backlight. The target signal consists of CII\,158 lines at $z\sim 3.6$, with lower redshift CO molecular transitions serving as interlopers. This diagram illustrates our simulations in individual slices, showcasing the geometric approach used to combine them into a single observed map that encompasses both the signal and interloping components. After generating simulated signal and interloper maps for all frequency channels of an experiment, this collection serves as a repository for conducting cross-correlations.}
\label{fig:int_visualization}
\end{figure*}

\subsection{Signal, Interloper, and Noise Modelling}
We generated the halo catalogues by conducting N-body simulations using the GADGET\footnote{\href{https://wwwmpa.mpa-garching.mpg.de/gadget4/}{https://wwwmpa.mpa-garching.mpg.de/gadget4/}} software package \citep{Gadget4}. Our simulation method involves creating 100 slices that spanned the redshift range from 20 to 0, each corresponding to an equal age of approximately 130 Myr, roughly the light travel time of the volume. The dimensions of the simulation box were set at 100 Mpc/$h$, and we achieved a length resolution of approximately 0.156 Mpc/$h$ (using a grid size of $N_{\rm grid}= 512$). This resolution enabled us to accurately resolve halo masses down to $10^{10}\, M_\odot/h$. 
The same random seed when generating snapshots for all 100 redshifts was used. The collective set of snapshots, evolving from a single Gaussian random initial condition while maintaining the same seed is referred to as \enquote{Sim-set 1}. Moreover, to capture variations along different lines of sight and account for the intrinsic properties of the Universe, we performed this process 13 times, altering the random seed each time. Consequently, we obtained 13 distinct sets of simulations(\enquote{Sim-set 1} to \enquote{Sim-set 13}), each evolving from different Gaussian random fields and offering unique representations of the Universe for different realizations. 

\begin{figure*}[t]
\includegraphics[width=\textwidth]{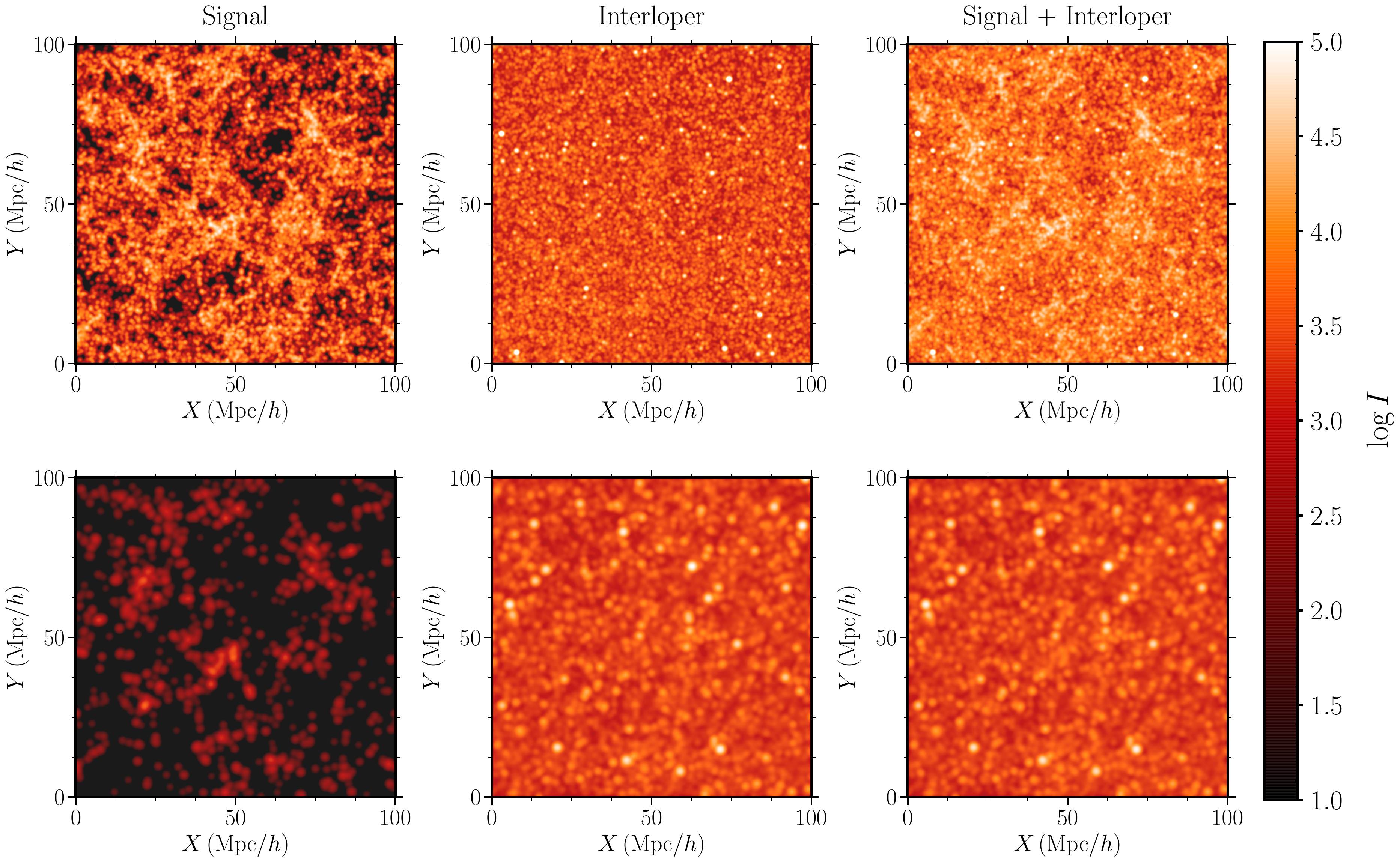} 
\caption{We showcase the combined map, which includes signals (CII\,158) at redshifts around $z\sim 3.7$ (top panels) and $z\sim 7.6$ (bottom panels), in addition to the maps of the signal and interlopers. The simulation covers a spatial range of 100 Mpc/$h$, with sources surpassing the minimum halo mass threshold of $10^{10} M_\odot/h$. The signal is simulated for the 410 GHz frequency channel, and the maps are smoothed with a 33 arcsec beam size. In the top panel, where the average signal strength is comparable to that of the interlopers, the combined map exhibits clear imprints of the signals. In contrast, in the lower panel, the combined map predominantly reflects the presence of interlopers due to the weaker signal strength.}
\label{fig:maps}
\end{figure*}

For each snapshot, we generate halo catalogues using the built-in friends-of-friends (FoF) algorithm with a linking length equal to 0.2 \citep{Gadget4}. From these halo catalogues, the next step is to create intensity maps by assigning different line intensities to the halos. This transformation of the halo catalogue into an intensity map is achieved through the utilization of the $\limpy$ package \citep{limpy}. This package takes the halo catalogue as input and produces an intensity map based on user-defined specifications. One of the key features of $\limpy$ is its ability to apply the beam convolution effect for a given experiment to the simulated intensity map. This effect simulates the effects of the telescope's beam size, which is characterized by the full width at half-maximum ($\Theta_{\rm FWHM}$) of a Gaussian beam. Additionally, $\limpy$ allows for the incorporation of frequency resolution ($\delta \nu$) when creating the 3D line intensity maps. Notably, the length resolution of the simulation box ($\Delta L_{\rm box}$) corresponding to the $N_{\rm grid}$ is not the same as the length resolution ($\delta L_{\nu}$) corresponding to $\delta \nu$. 

The $\limpy$ package offers a range of SFR and line luminosity models. These models can be employed to generate a set of line intensity maps based on specific approximations and assumptions. This flexibility allows us to create the desired signal at various redshifts, which is crucial for our observations using a telescope. Throughout this paper, we chose to employ the \enquote{Visbal10} model for studying the cross-correlation signal \citep{Visbal2010, limpy}. We have chosen this model due to its comprehensive calibration for all CO $J$-level transition lines, as well as the CII\,158 and OIII\,88 lines. This decision ensures consistency in our analysis, as we rely on this particular model to quantify both the signal and interloper contributions.

We assess the contribution of interloping signals by accounting for all the spectral lines that coincide within the same frequency channel as the desired signal in a given bandwidth of an experiment $\Delta \nu$. For example, consider observing CII\,158 emission from a redshift of approximately $z\sim 3.7$ at 410\,GHz. If we assume an experiment's bandwidth to be 40 GHz, several other lines, such as CO\,(4-3) at redshift $z\sim 0.12$, CO\,(5-4) at $z\sim 0.4$, CO\,(6-5) at $z\sim 0.68$, and less bright, higher J-level CO transitions, will act as interlopers in this frequency range. For each specific frequency channel, we calculate the redshift values corresponding to all the lines that could serve as interloping signals to the one we intend to observe. Subsequently, we create separate maps for each of these interloping lines originating at different redshifts than the signal. Then we compile a collection of maps, each representing a different line in total interloper contribution that shares the same frequency channel as our target signal.

To make a mock observation, we construct a lightcone from the simulated LIM snapshots that correspond to the frequency of interest, this includes the interloping sources. Notably, the interloping sources at lower redshifts exhibit larger sizes within the light-cone projection when compared to the target signal originating from higher redshifts. We note that two distinct interloping lines that originate from separate redshifts should not exhibit a correlation with each other. This lack of correlation is a consequence of the evolving galaxy properties and the growth of large-scale structures of the Universe between these two specific redshifts. As a result, when generating each simulated map for an interloping line within a particular simulation set, it is crucial to ensure that the maps of these two interloping lines are derived from entirely separate simulation sets. For instance, the map of one interloping line might be calculated from simulation denoted by \enquote{Sim-set 1}, while the other one is calculated from \enquote{Sim-set 2}. Once we have generated the lightcone simulation, that incorporates the target signal and all interlopers, we proceed to apply a beam convolution method. This method combines the maps of interlopers and signals to create a 3D representation of the Universe at a given frequency, helping us identify the unwanted signals in our observations. We carry out this same procedure for all six frequency channels, characterizing both the target lines from various LIM-LLX groups and the set of interloping signals that these frequency channels are set to capture.

\begin{figure*}[tb]
\includegraphics[width=\textwidth]{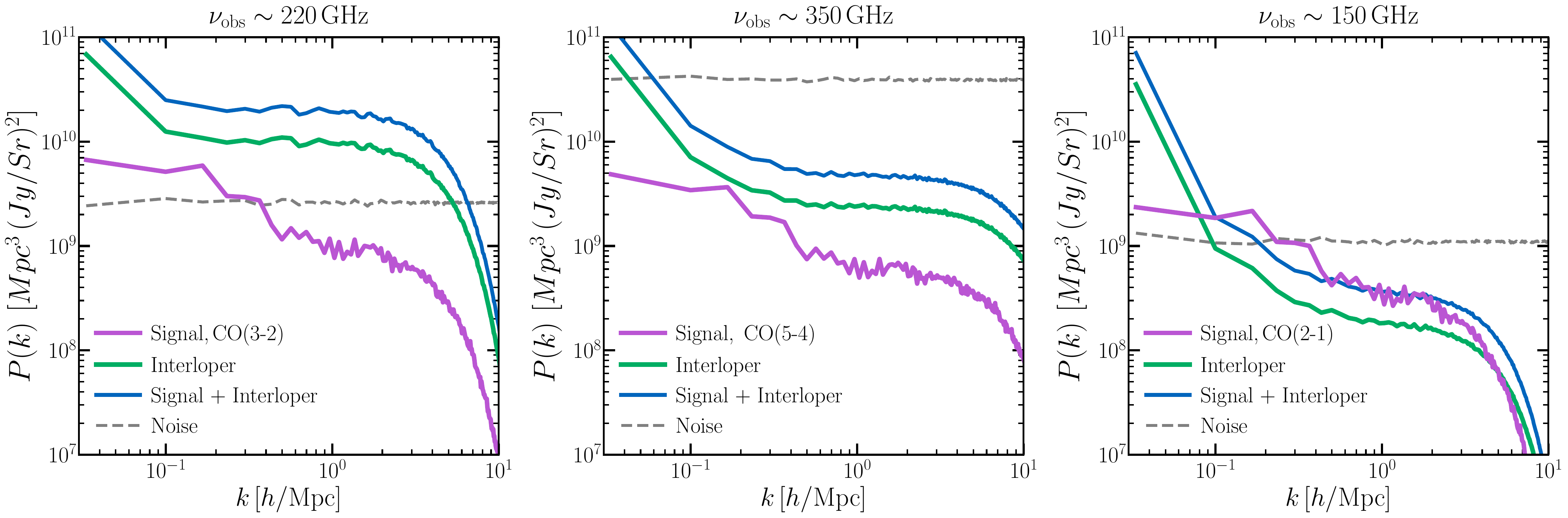} 
\caption{In the panels corresponding to the three distinct frequencies mentioned in the title, we illustrate the signal available for cross-correlation, instrumental noise, and interloper contamination. The mean redshift associated with the cross-correlation is approximately $z\sim 0.58$. Notably, the power spectra of the signal and interloper exhibit exponential decay at higher wave numbers ($k$) due to the beam-smearing effect.}
\label{fig:ps_separate}
\end{figure*}

Figure \ref{fig:int_visualization} provides a visual representation of signal and interloper contributions, illustrated through a schematic diagram. This figure serves as a visual summary of the method we follow to generate the lightcones. The lightcone presented here is specifically generated for the 410 GHz channel, where the primary source of interlopers consists of CO molecular lines. In contrast, the target line of interest is CII\,158 at a redshift of approximately $z\sim 3.7$. Moreover, we have overlaid a Cosmic Microwave Background (CMB) map in the background, providing a snapshot of the Universe at a redshift of approximately $z\sim 1100$. 

In figure \ref{fig:maps} we included projected maps of the interloper and CII\,158 emission for two distinct redshifts: $z\sim 3.7$ and $z\sim 7.6$. These maps correspond to observations made using the 410 GHz and 220 GHz frequency channels in the upper and lower panels, respectively. It is noteworthy that the average signal strength in the 410 GHz channel is approximately 55 times brighter than the signal simulated for the 220 GHz channel. In terms of interloper contamination, the 410 GHz channel exhibits interlopers that are 1.4 times brighter than those observed in the 220 GHz channel. When considering the total signal, which combines both the signal and interloper contributions, the 410 GHz channel is 2.6 times brighter than the same case observed at 220 GHz. Given the \enquote{Visbal10} model these cross-frequency comparisons of varying signal strengths and interloper contamination provide us with some intuition on how the following auto-spectrum and cross-spectrum results should scale with frequency.

\begin{table}[h]
\centering
\begin{tabular}{cccc}
\hline
\hline
$\nu_{\rm obs}$ & $\delta \nu_{\rm obs}$ & $\theta_{\rm FWHM}$ & $P_{N}$ \\
(GHz) & (GHz) & (arcsecond) & ($Mpc^3 Jy^2 sr^{-2})$\\
\hline
 90 & 0.9 & 78 & $9.2 \times 10^8$ \\
150 & 1.5 & 69 & $1.1 \times 10^9$ \\
220 & 2.2 & 58 & $2.6 \times 10^9$ \\
280 & 2.8 & 48 & $4.9 \times 10^9$ \\
350 & 3.5 & 37 & $3.9 \times 10^{10}$ \\
410 & 4.1 & 33 & $1.2 \times 10^{11}$ \\
\hline
\end{tabular}
\caption{The table provides information on the white noise power spectrum, beam size, and frequency resolution associated with each frequency channel. We assumed a total frequency bandwidth ($\Delta \nu)$ of 40\,GHz around the observational frequency channels.}
\label{tab:obs}
\end{table}

Finally, we incorporate instrumental noise into our analysis, which is necessary in order to make predictions for the detectability of auto and cross-correlated signals. In practical terms, two distinct sources of noise come into play: white random instrumental noise and atmospheric noise, both of which have the potential to influence the observation of line intensity maps. For the sake of simplicity, we did not include the effects of atmospheric noise in our analysis. In the future, we will have low-noise and high-resolution LIM experiments for the high-fidelity measurement of LIM signals. However, it is important to note that a reduction in noise does not alleviate the challenge posed by interlopers, since they will systematically bias the desired LIM signals and accounting for their contribution will depend on the detailed modelling of galaxy formation processes.

We generate a three-dimensional simulated noise under the Gaussian approximation. Initially, we created a simulation box with the same dimensions and grid points as those of the signal and interloper cubes. We populate the voxels within this box with random Gaussian numbers. Subsequently, we perform a Fourier transform on the box and normalize it using the white noise power spectrum value specific to the experiment, for our case, it is the EoR-Spec instrument on FYST. The noise realization in Fourier space can be mathematically expressed as:

\begin{equation}
N(k) = \sqrt{\frac{P_N}{V_{\rm cell}}} \widetilde{N}_{\rm box}.
\label{eq:Noise_fourier}
\end{equation}
\noindent In this equation, $\widetilde{N}(k)$ represents the three-dimensional noise realization in Fourier space, and $V_{\rm cell}$ is the volume of an individual voxel. $P_N$ is the white noise for a particular frequency channel as described in Table \ref{tab:obs}. Finally, we execute an inverse Fourier transform, capturing the real part of the result to generate the noise simulation in real space. This process enables us to model and incorporate noise characteristics in our simulations for any experiment.

\section{Quantifying Interlopers and Their Effects}\label{sec:quantifying}

We created lightcones for all the target signals, interlopers, and instrumental noise realizations, spanning across all six frequency channels as detailed in the previous sections. As illustrated in Figure \ref{fig:nu_z}, we carry out calculations for all feasible auto and cross-correlations involving the different frequency channels, where two lines originate from the same redshift range corresponding to the frequency bandwidth. However, we exclusively utilize the most prominent cross-correlations that could be probed by future LIM experiments (bolded in Table- \ref{tab:cross_groups}). These particular correlations serve as a test for evaluating the effectiveness of our interloper mitigation strategies. In all instances, we compute cross-correlations across various scenarios, including correlations solely between signals, those between signals with interlopers, and correlations encompassing signals, interlopers, and noise. This systematic approach enables us to gain insights into the levels of interloper contributions to various LIM signal detections.

In Figure \ref{fig:ps_separate}, we present specific signals of interest observed at three distinct observational frequencies, with the potential for cross-correlations with other frequency channels. Additionally, we provide an assessment of the extent of contamination stemming from both interlopers and instrumental noise. This Figure shows the amplitude of our primary signal of interest is not the dominant component of the overall total measurement. At the scale of $k\sim 1,h$/Mpc, we compare the variations in signal strengths among different lines probed by three different frequencies. The CO(3-2) signal at 220 GHz stands out, being nearly 1.5 times more prominent than the CO(5-4) signal at 350 GHz and approximately 2.8 times brighter than the CO(2-1) signal at 150 GHz. However, it is crucial to consider the influence of interloper contamination on the target signals. At 220 GHz and 350 GHz, interloper contamination is approximately 10 and 4 times more pronounced than the target signal at the same scale. In contrast, for the 150 GHz channel, the target signal CO(2-1) is brighter, surpassing the interloper contamination by a factor of 1.8. In this scenario, the detection of the CO(2-1) auto-power spectrum at 150 GHz may not be significantly impacted by interloper contamination but could instead be limited by the noise level. However, when we consider both white noise and interlopers, the total measured signal is nearly 14, 68, and 5 times greater than the target signals at 220 GHz, 350 GHz, and 150 GHz channels, respectively.

Therefore, in this scenario, directly detecting the auto-power spectrum would be challenging in the presence of both noise and interlopers. Hence, we proceed with the detection of the cross-power spectrum. This approach ensures that the noise and interloper contamination from two different frequency channels do not correlate, as they originate from different redshifts or different sources, for the interlopers (we assume the noise is uncorrelated between independent observations). We quantify how interlopers and noise can affect the signal-to-noise ratio for the detection of the auto and cross-power spectrum and its deviation from the target signal. We express the signal-to-noise ratio as 

\begin{equation}
    (S/N)^2 = \sum_i \frac{P_{\rm cross} (k_{i})^2} {\mathrm{Var}\,[P_{\rm cross} (k_{i})] },
\label{eq:snr}
\end{equation}
where $P_{\rm cross}$ represents the total measured power spectrum, and $i$ denotes the indices for the bins used in estimating the power spectrum. 

In addition to the signal-to-noise ratio, it is important to quantify whether the power spectra we measure are biased in the presence of interlopers, which we call the signal measurement bias. We chose to quantify this measurement bias using a statistic that is analogous to a reduced $\chi^2$, this statistic compares the measured total signal to the target signal calculated from the simulations. The expression for the measurement bias, denoted as $b_{\rm mes}$, is defined as follows:

\begin{equation}
b_{\rm mes} = \frac{1}{N_{\rm mes}} \sum_{i} \frac{[p^{i}_{\rm measured}(k) - p^{i}_{\rm target}(k)]^2}{\sigma_{i}(k)^2}.
\end{equation}
Here, $p^{i}_{\rm measured}(k)$ represents the measured cross-power spectrum, where the index $i$ signifies the specific bandpower under consideration. The target signal utilized in our analysis is denoted as $p^{i}_{\rm target}$ and $\sigma^{i}$ represents the error associated with $p^{i}_{\rm measured}$ within the context of cross-correlation. The total number of bandpowers in the measured power spectrum is denoted by $N_{\rm mes}$. This bias parameter serves as a crucial indicator of how the measured signal deviates from the input target signal, which is computed based on simulations and accounts for the statistical errors in each bandpower. The bias parameter approaches zero when the measured signal closely aligns with the target target signal or the errors are sufficiently larger than the signal. Conversely, when the biased parameter exceeds 1, it reveals a significant bias in the measured signal compared to the target signal. 

The volume of the simulation box is smaller than the survey volume of FYST's EoR-Spec-like experiment, therefore, we need to scale it with the appropriate volume factor when forecasting the SNR for that experiment. An experiment with a larger survey volume will measure a greater number of modes that are not present in the smaller simulation box. The observed number of modes within $\Delta k$ depends on the survey volume as $N_{\rm m}= k^2 \Delta k V_{\rm Survey}/ (2\pi)^2$, where $V_{\rm Survey}$ is the survey volume probed by an experiment. The variance or the error of the power spectrum decreases if the observed number of modes is increased, as $\sigma_i \propto N_{\rm m}^{-1/2}$. Hence, the error bars for a particular bin are proportional to the survey volume. Hence, we plot the signal-to-noise ratio for the simulation box, but these can be scaled by a factor of $\sqrt{V_{\rm Survey}/V_{\rm box}}$. We note that this scaling does not account for any variance in the initial SNR calculation on the simulation box, which is subject to fluctuations from cosmic variance.

\begin{figure}[h]
\includegraphics[width=0.5\textwidth]{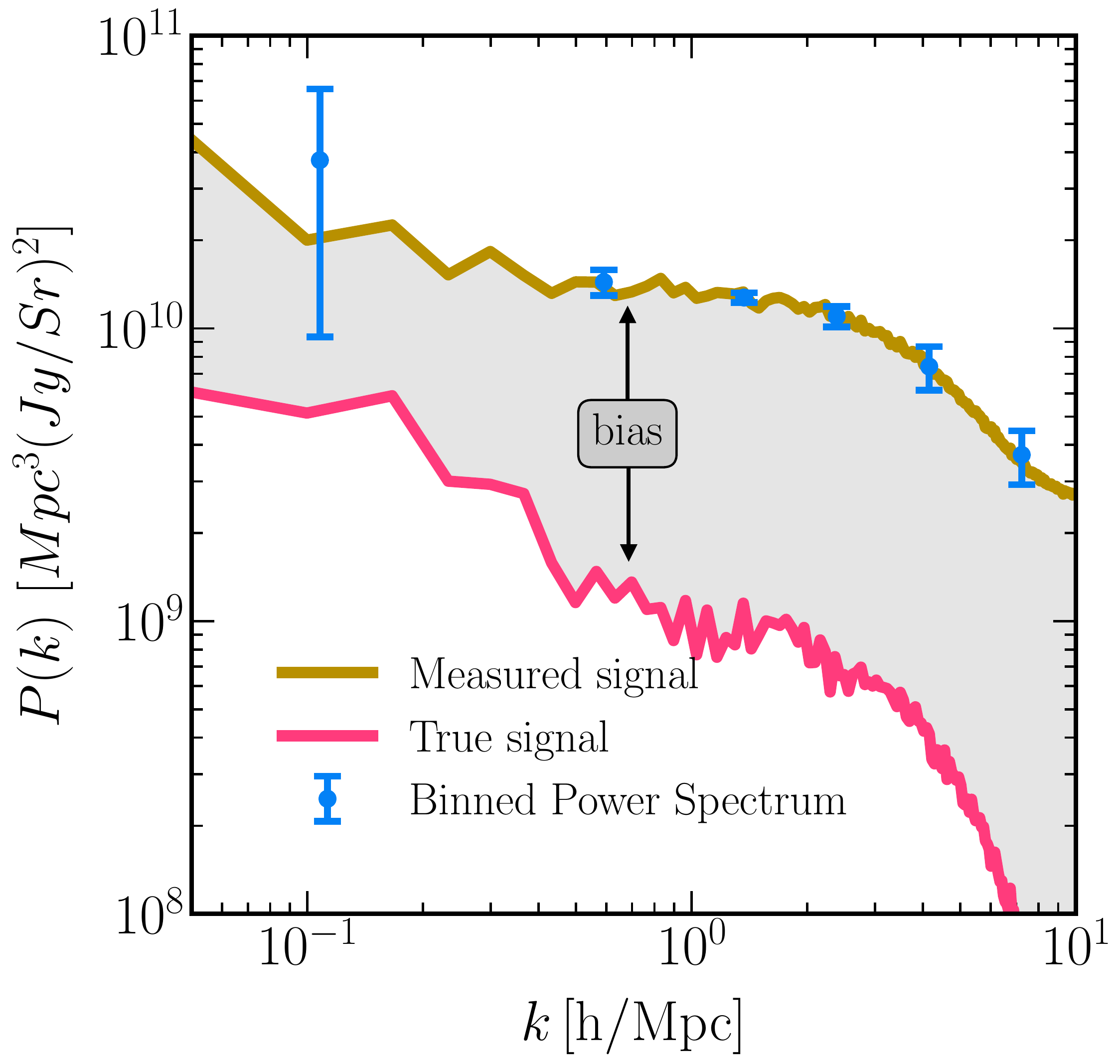} 
\caption{Forecasts for the measured auto power spectrum of the CII\,158 line intensity map at approximately $z\sim 3.7$ are generated using the 410 GHz frequency channel. We display corresponding error bars along with the target signal estimated from the map's variance for comparison. Despite the small error bars, they do not track the target signal, indicating a significant measurement bias. This discrepancy arises because we effectively detect the combined contributions of interlopers and instrumental noise, which are brighter than the target signals.}
\label{fig:ps_auto}
\end{figure}

The survey volume of an experiment can be expressed in terms of the survey area $S_A$ and the frequency bandwidth $B_\nu$ as \citep{Dumitru2018}

\begin{multline}
V_{\rm surv}= 3.7 \times 10^{7} (\mathrm{cMpc}/h)^3 \left(\frac{\lambda_{\rm line}}{157.8\,\mu m}\right) \left(\frac{1+z}{8}\right)^{\frac{1}{2}} \\
\times \left(\frac{S_A}{16\, 
\rm {deg^2}} \right) \left(\frac{B_\nu}{20\, \mathrm{GHz}} \right)\,.
\label{eq:Vsurv}
\end{multline}
Here $\lambda_{\rm line}$ represents the rest frame wavelength of the emitted lines, $S_A$ stands for the effective survey area of the experiment, and $B_{\nu}$ denotes the frequency bandwidth. In the context of an experiment akin to EoR-Spec conducted on FYST, we adopt a constant value of $B_{\nu} = 40$ GHz across all frequency channels. Now, we can scale the SNR estimated primarily from the simulation box for an experiment.

\subsection{Impact on the Auto-power Spectrum}
Detecting the auto-power spectrum of multiple spectral lines spanning a broad range of redshifts is of great importance for gaining insights into the characteristics of galaxies and the underlying cosmic structures. In the context of both first and second-generation LIM experiments, the presence of interloper contaminants, in conjunction with the inherent noise, constitutes a formidable hurdle to achieving unbiased detections. This challenge becomes even more pronounced when dealing with observations of the CII\,158 lines at high redshifts, where the presence of various bright J-level CO transitions can manifest as a significant foreground signal.

In Figure \ref{fig:ps_auto}, we show the impact of interlopers on the detection of the auto power spectrum of CII\,158 lines at a redshift of approximately $z\sim 3.7$. Our method for estimating error bars on the binned power spectrum entails calculating the variance within a bin centered at $k_{\rm cen}$ with a width of $\Delta k$. The signal-to-noise ratio (SNR) is computed to be 294, while the bias parameter is measured at 8.3, so the measured signal in each bandpower is on average $\sim8\sigma$ away from the underlying CII signal. This amount of bias indicates that the SNR detection is not purely indicative of the target  signal; instead, our observation comprises a combination of interloper contamination and instrumental noise. The example here clearly demonstrates the importance of any efforts to develop strategies that successfully reduce the impact of interlopers in LIM studies.

The mitigation of noise contribution in the auto power spectrum has been extensively investigated across various fields, including CMB analyses, where methodologies such as data splitting are commonly employed \citep[e.g.,][]{ACT2020choi}. In this approach, the dataset corresponding to each set of maps is partitioned into two or more independent splits, followed by cross-correlation of these splits to attenuate noise bias. It is anticipated that there will be a decorrelation in instrumental noise due to variations in data collection times or seasons, while the signal should exhibit correlation at the map level. The efficacy of this technique has been thoroughly examined in the context of LIM experiments, such as MeerKAT and COMAP \citep{COMAP-IV, Meerkat-detection}. Throughout this work when we refer to noise, we are specifically referring to instrumental noise and not noise bias. Beyond this subsection, we turn the focus of this work to cross-correlations, where the noise properties between different frequency maps are assumed to be uncorrelated and the noise bias will not be present.

For LIM it is essential to employ a suitable strategy for characterizing and removing interlopers to obtain unbiased auto-power spectra at different redshifts. One potential approach in this endeavor involves eliminating the interloper contributions from the overall observed map, followed by a comprehensive map-level analysis \citep{Lidz2016-interloper}. However, this method requires careful modeling of what fraction of the interloping sources remain, which can exhibit variability depending on the templates used. Alternatively, a second approach entails the application of Principal Component Analysis (PCA) at the power spectrum level \citep{Zhou:2022pca, Concerto-ps}. 

\begin{figure}[h]
\includegraphics[width=0.5\textwidth]{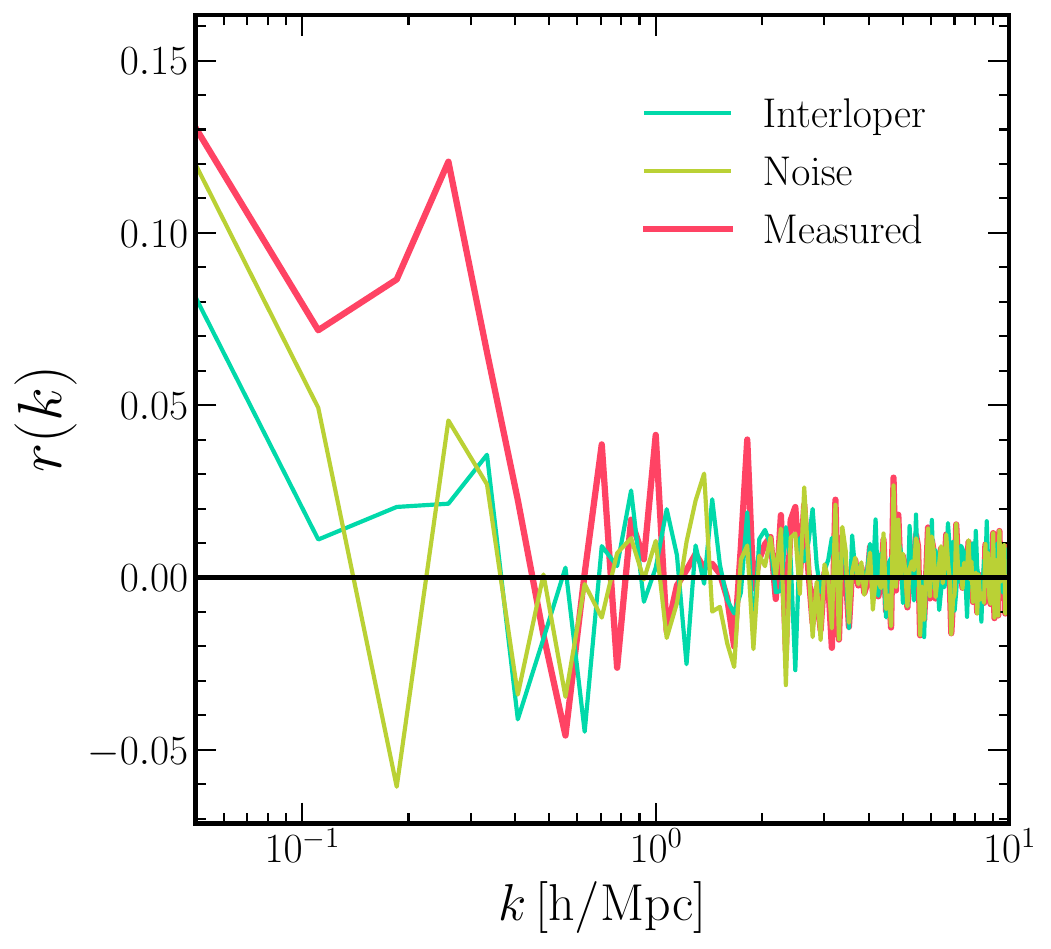} 
\caption{
The cross-correlation coefficients when we cross-correlate maps simulated for two distinct frequency channels, for interlopers, noise, and a combined map that includes signal, noise, and interlopers. Incorporating the signal into the combined map of noise and interlopers boosts cross-correlation coefficients at larger scales, as both frequency channels share correlated signals from the same sources. At small scales, the signal has a limited impact on cross-correlation coefficients as its strength diminishes exponentially due to beam convolution.}
\label{fig:r_c4}
\end{figure}

\begin{figure*}[t]
\includegraphics[width=\textwidth]{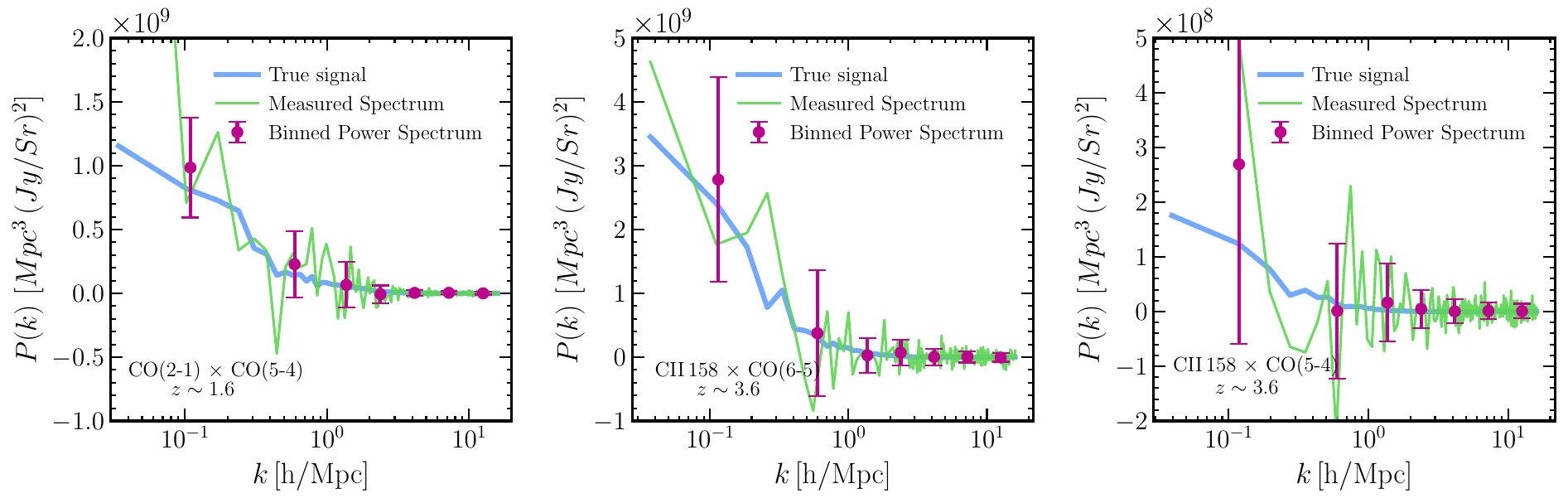} 
\caption{We demonstrate the feasibility of detecting LIM-LLX even in the presence of interlopers. The left, center, and right panels depict forecasted error bars for cross-correlation groups C2, C4, and C6, respectively. These error bars encompass both the target signal and the measured signal, accounting for residual noise and interloper contributions. Please note that the error bars in these specific simulations will reduce by a factor of ($\sqrt{V_{\rm S}/V_{\rm box}}$) as the number of modes increases for a larger box size. Despite their lower SNRs (see Table~\ref{tab:bias_SNR}), the LIM-LLX combinations in this figure do not exhibit biases in their measurements.}
\label{fig:cross_error}
\end{figure*}

\subsection{Impact on the Cross-power Spectrum}
We examine the correlations among signals, noise, and interlopers, as well as their various combinations, between two distinct frequency channels. If noise and interlopers exhibit strong correlations between these channels, it can lead to biases in the measured cross-correlation. To assess the effectiveness of our analysis technique, we introduce a parameter to quantify cross-correlation efficiency, denoted as $r(k)$. This parameter is defined by the following expression:

\begin{equation}
r(k) = \frac{P^{X}{\nu_1 \nu_2} (k)}{\sqrt{P^{X}_{\nu_1} (k) \times P^{X}_{\nu_2} (k)}}.
\label{eq:r}
\end{equation}
Here $P^{X}_{\nu_1 \nu_2} (k)$ represents the cross-power between two distinct spectral lines observed at frequencies $\nu_1$ and $\nu_2$. The symbol \enquote{X} denotes the cross-power spectrum under various scenarios, including signal-only, signal and interloper, and the total measured signal, including signal, instrumental noise, and interloper contributions. The terms $P^{X}_{\nu_1}$ and $P^{X}_{\nu_2}$ are the auto power spectra of two specific lines of interest observed at distinct frequencies.

In Figure \ref{fig:r_c4}, we show the correlations between interloper, signal, and the total measured signal at a redshift of approximately $z\sim 3.7$. We use the maps for all the interlopers, signal, and noise components at both 410 GHz and 150 GHz frequencies. The focus here centers on the detection of the two brightest target lines within the C4 group, namely, CII\,158 and CO(6-5). Looking at $r(k)$ on small scales, specifically for wave numbers $k \gtrsim 1$ $h$/Mpc, a trend emerges. The average correlation between noise components from two distinct frequency channels exhibits fluctuations around zero. Similarly, the average cross-correlation efficiency tends towards zero when interlopers are considered separately. However, when we introduce the signal into the maps this is no longer the case. The correlation function increases for the total signal, which includes, the signal, noise, and interloper components in the maps. This observation underscores the fact that signals emanating from two separate frequency channels exhibit a degree of correlation, even in the presence of interloping sources and noise within the map. However, it's essential to note that at small scales, the behavior of the total signal closely resembles that of the interloper and noise maps. This result is attributed to the significant damping effect induced by the beam on the signal at such scales. Since the beam-convoluted signal strength is significantly lower than the total measured signal, which includes both noise and interloper contributions, the amplitude of these small-scale fluctuations is not large enough to show a significant correlation over the fluctuations from noise and interlopers.

\begin{table}[tb]
\centering
\begin{tabular}{lcc|cc|c}
\hline\hline
 & \multicolumn{2}{c}{Total}
& \multicolumn{2}{c}{Interlopers} &\multicolumn{1}{c}{Total} \\
\cline{2-3} \cline{4-5} \cline{5-6} 
Groups  & SNR & $b_{\rm mes}$ & SNR & $b_{\rm mes}$ & SNR (FYST) \\
\hline
  C1 & 3.7 & 0.20 & 10 & 0.34 & 34 \\
  C2 & 2.7 & 0.11 & 5.0 & 0.15 & 23 \\
  C3 & 1.2 & 0.06 & 3.8 & 0.30 & 14 \\
  C4 & 1.8 & 0.06 & 2.9 & 0.13 & 10 \\
  C5 & 3.4 & 0.27 & 4.5 & 0.30 & 26 \\
  C6 & 0.87 & 0.03 & 2.1 & 0.08 & 5 \\
  C7 & 0.83 & 0.12 & 1.8 & 0.23 & 5 \\
\hline
\end{tabular}
\caption{We provide quotations for the forecasted SNR and measurement bias parameter across multiple cross-correlation groups. The term 'Total' denotes the SNR, considering contributions from both instrumental noise and interlopes. In the last column, we scale the SNR for an EoR-Spec-like experiment on FYST by scaling the SNR by the survey volume factor (see Equation~\ref{eq:Vsurv}).}
\label{tab:bias_SNR}
\end{table}
\newpage
\section{Detectability of LIM-LLX}\label{sec:detectability}
In this section, our objective is to forecast the potential detectability of LIM-LLX between various lines belonging to the cross-correlation groups C1 to C7, utilizing the experimental setup similar to FYST's EoR-Spec-like experiment. To assess the feasibility of mitigating the impact of interlopers and noise through LIM-LLX, we use metrics described in the previous section: the SNR and the bias parameter. Our analysis encompasses all cross-correlation groups featuring the brightest lines. Furthermore, we incorporate additional low-frequency channels at 90 and 150 GHz, along with the EoR-Spec's planned frequency coverage. With these additional channels we address the question: what proves to be more effective in interloper mitigation; deploying more detectors at higher frequencies, which reduces the instrumental noise level; or deploying new low-frequency channels, which increases the wavelength coverage?
Our analyses here provide insights into optimizing experimental setups that will enhance the measurement fidelity by future LIM experiments.

In Figure \ref{fig:cross_error}, we show the simulation results for LIM-LLX of the C2, C4, and C6 cross-correlation line groups. These results include the input signal, the simulated total cross-power spectrum, which serves as a substitute for what an EoR-Spec-like experiment might measure, and the associated error bars on the power spectrum bandpowers. It is important to emphasize that these error estimates for the cross-power spectrum bandpowers are calculated from the simulated volume, which is characterized by $V_{\rm box}=100^3$ (Mpc/$h)^3$. Therefore, the errors shown in this figure are larger than that of the full projected survey from an EoR-Spec-like experiment by the ratio between the simulation box size and the survey volume, $\sqrt{V_{\rm box}/ V_{S}}$. An experiment conducted within a vast survey volume inherently encompasses a greater number of modes that extend beyond the confines of our simulation. This expanded survey volume grants access to an array of larger-scale modes that are beyond the reach of our simulated data. Therefore, we simultaneously calculate the SNR based on the simulation box and then scale it to match the conditions of an EoR-Spec-like experiment. This approach allows us to account for the differences in survey volume and accurately assess the SNR in a real observational setting.

\begin{figure}[t]
\centering
\includegraphics[width=0.48\textwidth]{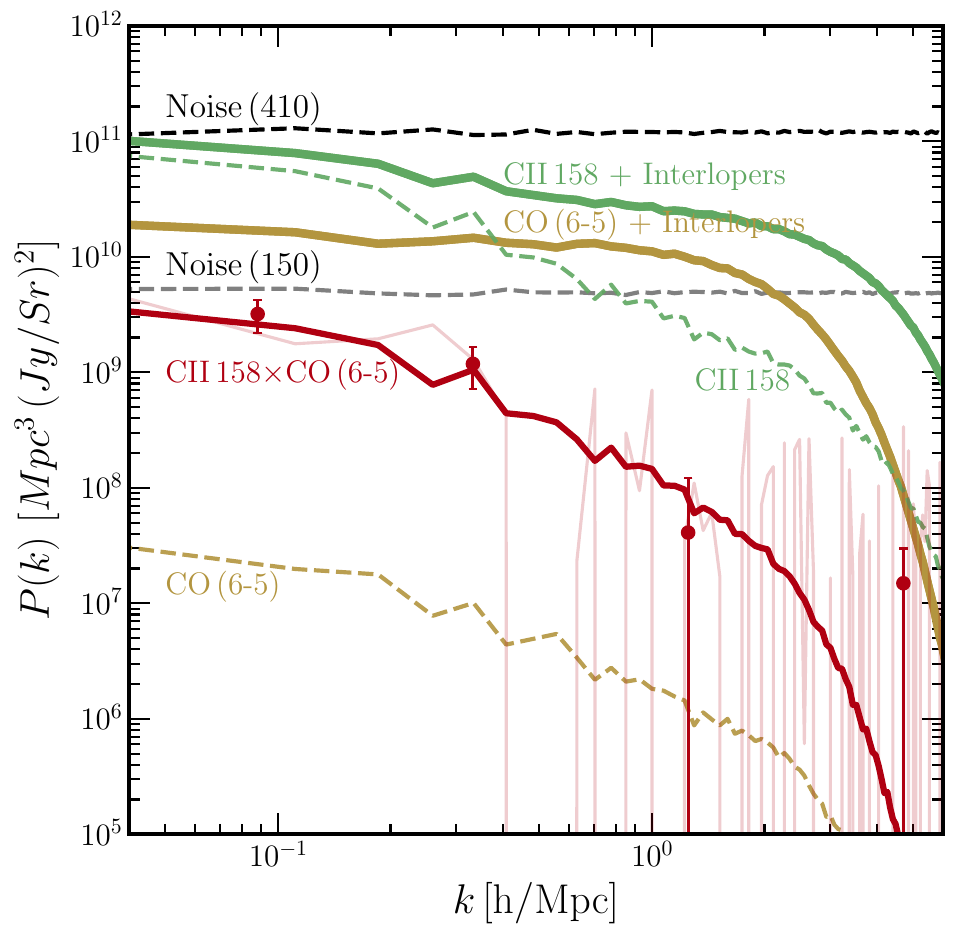} 
\caption{The figure illustrates the cross-power spectrum of CII\,158 and CO(6-5) lines at redshift $z\sim 3.7$. Additionally, it presents the total auto power spectra of CII\,158 and CO(6-5) lines at the same redshift, taking into account instrumental noise and interlopers as annotated in the figure. The light red shade represents the measured cross-power spectrum from simulated noise-dominated data of CII\,158 and CO(6-5), while the solid red line depicts the actual cross-correlated signal. We include error bars after binning the measured power spectrum which are consistent with the target signal. This highlights the substantial reduction in bias achieved through cross-correlation between the two frequency channels (target lines), as demonstrated in the figure.}
\label{fig:summary_plot}
\end{figure}

Table \ref{tab:bias_SNR} provides an overview of the SNR values and the signal bias parameter, our chosen metrics for assessing the ability to detect unbiased LIM-LLX signals. We calculate the SNR in two ways: 1) when forecasting the detectability of the target signal, considering the bias from both instrumental noise and interlopers, denoted by the term "Total", and 2) when forecasting the same while considering the bias caused by only interlopers, with noise bias being negligible.
We estimate that the highest LIM-LLX SNR achievable by FYST's EoR-Spec is 34, which can be attained through the cross-correlation between CO(3-2) and CO(5-4) lines at a redshift of approximately $z\sim 0.58$. This particular correlation allows for the probing of galaxies at low redshift with high significance. The increase in SNR between the simulation calculation and the forecasted FYST's EoR-Spec measurements arises from the fact that the survey volume in this scenario is significantly larger, approximately 271 and 76 times greater than the volume of the simulation box, at $z\sim 0.58$ and $7.6$, respectively. Additionally, we have identified another notable SNR value of 25, pertaining to the cross-correlation between CII\,158 and CO(4-3). This SNR level is sufficient to study the post-reionization physics of galaxy formation through LIM observations. Our analysis reveals a modest SNR of 5.1 for the detection of LIM-LLX between CII\,158 and OIII\,88 at $z\sim 7.6$. Despite its relatively lower value, this SNR holds significant promise for advancing our understanding of early galaxies and their crucial role in the Universe's reionization. This redshift roughly marks a pivotal phase during the epoch of reionization which, according to some current models, nearly half of the Universe was already re-ionized. 

We find that, in all seven cross-correlation groups, there is a substantial reduction in the bias parameter. This reduction consistently brings the bias parameter to values below 1, so the bias is well below the statistical error. This outcome signifies an unbiased measurement of cross-correlation, underscoring the effectiveness of our analysis techniques. This is in stark contrast to auto power spectrum measurements, where signal bias remains significant in the presence of interlopers. Therefore, a FYST-like experiment spanning the frequency channels from 90 to 410 GHz possesses the capability to conduct a comprehensive, tomographic exploration across redshifts ranging from 0.58 to 7.6, achieved through the cross-correlation of various spectral lines, enhancing our ability to probe the Universe across different epochs.

In Figure \ref{fig:summary_plot}, we present a summary of our results that underscore the effectiveness of cross-correlation techniques. As previously detailed, the inclusion of target signals in the cross-correlation, specifically between the combined map of noise and interlopers, significantly enhances the fidelity of cross-correlation at larger scales. We find that the measured cross-correlation signal accurately follows the target simulated signal at these larger scales, resulting in minimal errors in the cross-spectrum. However, at smaller scales, our observations primarily reflect the cross-correlation between noise and interlopers, as the signal's impact is limited due to the beam-smearing effect. Hence, the measured signal exhibits large fluctuations, leading to larger error bars. Unlike the auto power spectrum, we note that these error bars remain consistent with the target signal and effectively capture the scale-dependent behavior of the cross-power spectrum.

\begin{figure}[h]
\centering
\includegraphics[width=0.48\textwidth]{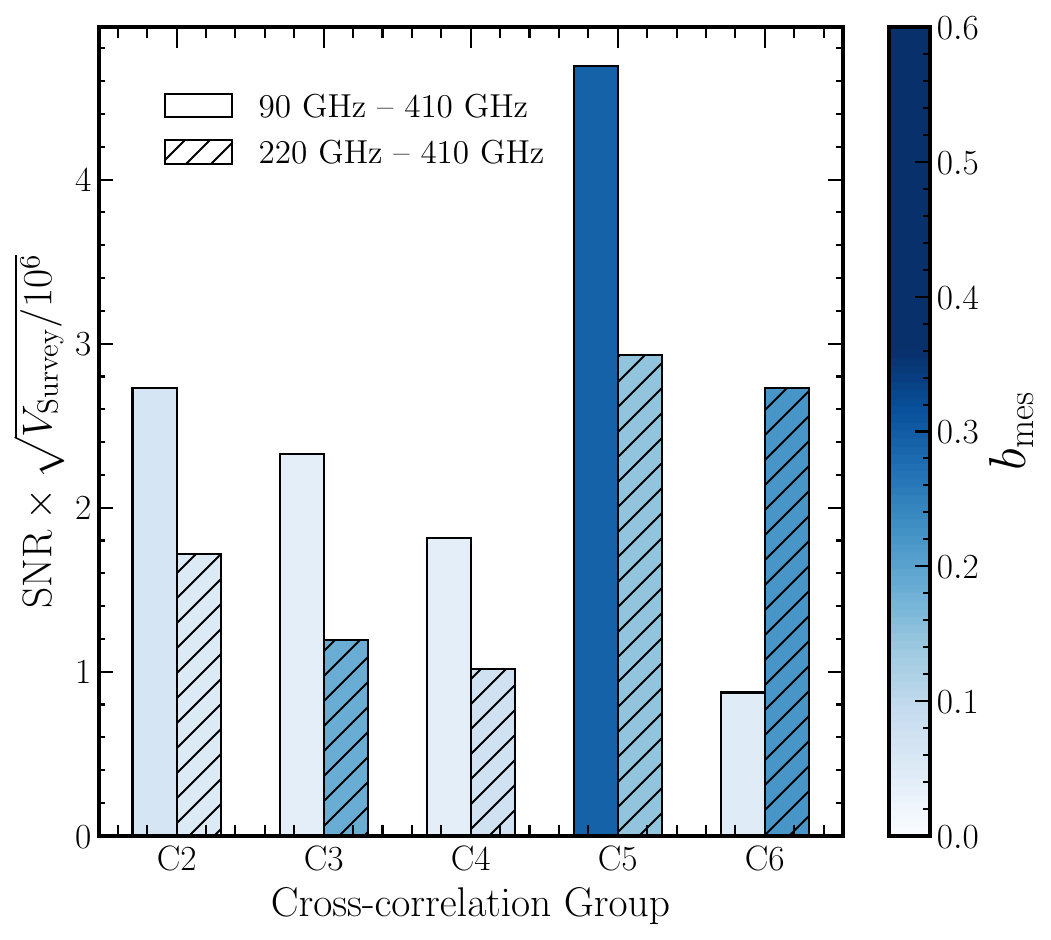} 
\caption{This figure illustrates the relationship between the bias parameter and the signal-to-noise ratio (SNR) across five cross-correlation groups. Bars without stripes represent the bias parameter values for cross-correlation groups that incorporate two additional frequencies, namely 90 GHz and 150 GHz, taking into account both residual noise and interloper contributions. In contrast, striped bars indicate the parameters for the LIM-LLX group with prime notation, which utilizes EoR-Spec's original four frequency channels, considering only the interloper contribution (no noise). This figure shows that incorporating low-frequency channels generally improves SNR for LIM-LLX in most cases over reducing the noise of existing channels, except for the C6 group, while also mitigating measurement bias for most groups, except for C5.}
\label{fig:snr_bias}
\end{figure}

With Figure \ref{fig:snr_bias} we address the question of whether the addition of more low-frequency channels or increasing instrument sensitivities across existing frequency channels proves to be more effective in interloper mitigation. This figure shows the SNR for the correlation groups C2 through C6 with the color signifying the bias, with the goal of highlighting which LIM-LLX groups benefit from broader frequency coverage (solid bars) or lower noise (dashed bars).
For the lower noise, we consider only the interloper contamination. This choice assumes the maximal case where a sufficient number of detectors are deployed such that the instrumental noise is significantly small in comparison to the interloper contribution. 

The correlation groups denoted C1 through C7 included the enhanced frequency coverage, while the correlation groups denoted C2$^\prime$ through C6$^\prime$ represent the lower noise scenario. We find that the LIM-LLX associated with the C2 group exhibits an SNR of 2.7 with a corresponding bias parameter of 0.11. If we consider solely the interloper contributions for the C2$^\prime$ group, excluding noise effects, the SNR notably decreases to 1.7. This observation suggests that by reducing noise at high-frequency channels, we do not achieve any improvements compared to using the low-frequency channel. This is because, with one additional 90 GHz channel, we can perform cross-correlations between brighter lines, such as CO(2-1) and CO(5-4). However, if we exclude the 90 GHz channel, we have the alternative option to cross-correlate CO(5-4) and CO(8-7). In this case, CO(8-7) is considerably fainter than CO(5-4), resulting in a lower value of SNR even though the overall instrumental noise level is reduced. Similarly, we compare the SNR for prime LIM-LLX groups against the non-prime LIM-LLX groups and we find that we do not gain in SNR except the group C6$^\prime$. Here, the reduction of instrumental noise outweighs the decrease of cross-correlation signal from dimmer CO lines. This figure underscores the benefits of adding low-frequency channels for probing LIM-LLX across a wide redshift range. The addition of low-frequency channels typically enhances SNR and reduces bias, leading to more precise measurement.

\section{Discussion and Conclusion}\label{sec:conclusion}
Cross-correlation techniques between two separate spectral lines sourced from the same origins are of paramount importance in the field of astrophysics. In cross-correlation, the LIM-LLX technique provides observational probes that are highly valuable for the detection of target astrophysical signals. One of LIM-LLX's most significant contributions is its ability to interpret the correlation in complex data, probe genuine astrophysical signals and mitigate unwanted interlopers and instrumental noise. This capability could play a crucial role in LIM experiments, where precision and reliability are very important for estimating the astrophysical and cosmological parameters from LIM observations. 

In this work, we present a simulation-based study to examine the reliability and effectiveness of LIM-LLX amidst the presence of interlopers and simulated noise. Our primary objective is to emulate the intricacies of real LIM observations within a simulated volume spanning 100 Mpc/$h$. Within this framework, we explored the potential cross-correlation groups, denoted in prime notation, which can be executed using FYST's EoR-Spec frequency range spanning from 220 to 410 GHz. This extensive range of frequencies allows us to conduct cross-correlations not only with the CII\,158 lines but also with higher J-level CO transitions, even up to redshifts of $z\sim 3.7$. Additionally, our investigation delves into the feasibility of extending LIM-LLX to explore the low redshift Universe. To achieve this, we introduce two extra frequency channels, operating at 90 and 150 GHz, enabling us to probe the intricacies of cross-correlations between different CO J-level transition lines. Our analysis illustrated the potential and limitations of LIM-LLX as a tool to advance our understanding of astrophysical phenomena across a wide range of redshifts. 

We conducted a analysis of the primary sources of contamination that could impact the detection of specific line emissions. We identify two primary sources of contamination for the LIM observations: instrumental noise and interloper contributions. We discuss the potential impact of these factors on our results and elaborate on strategies to mitigate their effects on LIM-LLX.

As an example, we calculated of the auto power spectrum for CII\,158, revealing a notably skewed outcome primarily attributed to the substantial influence of interlopers and instrumental noise. This skewed result yielded a systematic bias $\sim 8$, signifying a considerable level of contamination. To mitigate this bias and enhance the reliability of our measurements, we employed a cross-correlation approach between CII\,158 and CO(6-5), utilizing both the 410 GHz and 150 GHz frequency channels. This strategic combination yielded a remarkable reduction in the bias parameter to 0.06. Such a substantial reduction underscores the effectiveness of cross-correlation techniques in isolating and amplifying genuine astrophysical signals while minimizing the impact of unwanted interlopers and noise. However, we note that this does come at the cost of a reduction in the overall SNR. This particular case study serves as a compelling exemplar, illustrating the capacity of LIM-LLX to increase the fidelity of intensity mapping measurements, even in the challenging presence of prevalent contaminants. We found that incorporating low-frequency channels like 90 GHz and 150 GHz into an EoR-Spec-like experiment is more crucial for measuring unbiased cross-correlated signals across a wide redshift range than reducing the noise levels in the already existing high-frequency channels. This is because low-frequency channels allow us to probe the molecular lines from CO at low redshifts, which are considerably brighter than the higher-order J-level CO transitions observed by high-frequency channels. 

An additional challenge in signal detection arises from continuum emission, which can introduce bias into our measurements. To address this issue, one can adopt the Principal Component Analysis (PCA) method, as previously suggested by \citet[]{Concerto-ps}. By applying PCA, one can effectively suppress the bias caused by continuum emission, thus enhancing the fidelity of the LIM signals. We leave the mitigation of continuum emission to future work. In this paper, we employ a simple model to create signal and interloper maps. We did not consider the scatter relation around the mean line luminosity and SFR (or halo mass) relation. Consequently, in the absence of interlopers and instrumental noise, the cross-correlation coefficient between two target lines at redshift $z$ theoretically reaches 1. However, this may not hold true if we incorporate scatter and stochasticity when modeling the line luminosities for two target lines at the same redshift. To address this, we intend to incorporate the scatter relation into the $\limpy$ package and assess the consistency of the findings presented in this study.

While our analysis does not incorporate atmospheric noise, it is important to recognize its potential influence in future studies. Accounting for atmospheric noise will be crucial in refining the accuracy of LIM experiments \citep{CCAT-prime2021}. In future work, we will incorporate atmospheric noise into our analytical framework. This addition will enable us to assess the influence of atmospheric noise on both the SNR and bias parameters, shedding light on how these critical metrics evolve in the presence of atmospheric factors. Furthermore, we plan to integrate this atmospheric noise component into a specific experiment's scanning strategy. This strategic integration will pave the way for a map-level analysis. By analyzing maps extracted from observational data obtained during the experiment, we aim to perform precise parameter estimations. These estimations will yield unbiased measurements of the cosmological and astrophysical parameters of our interest.

\section{Acknowledgement}
We thank Patrick Breysse, Steve Choi, Dongwoo Chung, Abby Crites, Michael Niemack, and Anthony Pullen for helpful discussions. AR is partially supported by the CCAT collaboration and also acknowledges support from NASA under award number 80NSSC18K1014 during the final stage of this work. NB acknowledges
support from CCAT collaboration for this work and acknowledges additional support from NASA grants 80NSSC18K0695 and 80NSSC22K0410.

\bibliographystyle{aasjournal}
\bibliography{citation}


\end{document}